\documentclass[numberedappendix,natbib209]{emulateapj}

\usepackage{epsfig}
\usepackage{longtable}
\usepackage{natbib}

\newcommand{\cii}{[C\,{\sc ii}]}

\newcommand{\mgii}{Mg\,{\sc ii}}

\newcommand{\lya}{Ly$\alpha$}

\newcommand{\myemail}{chris.willott@nrc.ca}

\def\co21{CO\,(2-1)}

\shorttitle{Redshift 6.4 host galaxies of $10^{8}$ solar mass black holes}
\shortauthors{Willott et al.}

\begin{document}


\title{Redshift 6.4 host galaxies of $10^{8}$ solar mass black holes:\\ low star formation rate and dynamical mass}


\author{Chris J. Willott}
\affil{Herzberg Institute of Astrophysics, National Research Council, 5071 West Saanich Rd, Victoria, BC V9E 2E7, Canada}
\email{\myemail}

\author{Alain Omont and Jacqueline Bergeron}
\affil{UPMC Univ Paris 06 and CNRS, UMR7095, Institut d'Astrophysique de Paris, F-75014, Paris, France}


%\altaffiltext{1}{Visiting Astronomer, Cerro Tololo Inter-American Observatory.
%CTIO is operated by AURA, Inc.\ under contract to the National Science


\begin{abstract}
  We present ALMA observations of rest-frame far-infrared continuum
  and \cii\ line emission in two $z=6.4$ quasars with black hole
  masses of $\approx 10^8 M_\odot$.  CFHQS\,J0210-0456 is detected in
  the continuum with a 1.2\,mm flux of $120\pm 35\,\mu$Jy, whereas
  CFHQS\,J2329-0301 is undetected at a similar noise level. J2329-0301
  has a star formation rate limit of $<40\,{\rm M}_\odot\,{\rm
    yr}^{-1}$, considerably below the typical value at all redshifts
  for this bolometric luminosity. By comparison with hydro
  simulations, we speculate that this quasar is observed at a
  relatively rare phase where quasar feedback has effectively shut
  down star formation in the host galaxy. \cii\ emission is also
  detected only in J0210-0456. The ratio of \cii\ to far-infrared
  luminosity is similar to that of low redshift galaxies of comparable
  luminosity, suggesting the previous finding of an offset in the
  relationships between this ratio and far-infrared luminosity at low-
  and high-redshift may be partially due to a selection effect due to
  the limited sensitivity of previous continuum data. The \cii\ line
  of J0210-0456 is relatively narrow (FWHM\,$=189 \pm
  18$\,km\,s$^{-1}$), indicating a dynamical mass substantially lower
  than expected from the local black hole -- velocity dispersion correlation. The
  \cii\ line is marginally resolved at $0\farcs7$ resolution with the
  blue and red wings spatially offset by $0\farcs5$ (3\,kpc) and a
  smooth velocity gradient of $100$\,km\,s$^{-1}$ across a scale of
  6\,kpc, possibly due to rotation of a galaxy-wide disk. These
  observations are consistent with the idea that stellar mass growth
  lags black hole accretion for quasars at this epoch with respect to
  more recent times.

\end{abstract}





\section{Introduction}

The peak of global star formation occurred about 10 billion years ago
at a redshift of $z\approx 2$ (Reddy \& Steidel 2009). The rise in
star formation at earlier times is studied by tracing the space
density and properties of higher redshift galaxies. Such galaxies can
be selected in the rest-frame ultraviolet as Lyman break dropouts
(Bouwens et al. 2006), in the rest-frame infrared as line or continuum
sources (Carilli \& Walter 2013), via black hole accretion activity as
active galactic nuclei (AGN) or quasars (Fan et al. 2006), or via
stellar explosions such as gamma ray bursts (Tanvir et
al. 2012). These methods are complementary in that they are sensitive
to galaxies with a range of mass, star formation rate, dust formation
rate and black hole accretion rate, to allow a broad view of early
galaxy evolution.

Thanks to the high fraction of stellar radiation re-radiated in the
infrared by interstellar dust and gas and the negative k-correction
(Blain \& Longair 1993), high redshift galaxies can be well studied by
millimeter observations. Continuum observations are sensitive to the
star formation rate as radiation from young, hot stars is re-radiated
by dust. Molecular lines such as CO probe the molecular gas in dense
star-forming regions. Atomic lines such as the fine-structure line of
singly-ionized carbon, \cii, probe the interstellar medium and the
outer parts of star-forming regions. It has been recognized that the
\cii\ line will likely become the most useful line for studying very
high redshift galaxies with the Atacama Large Millimeter Array (ALMA;
Walter \& Carilli 2008). Indeed early ALMA observations already show
detections of \cii\ in normal star-forming galaxies at $z>4$ (Wagg et
al. 2012; Carilli et al. 2013). \cii\ has also been detected in a few
$z>6$ quasar host galaxies (Maiolino et al. 2005; Walter et al. 2009;
Venemans et al. 2012; Wang et al. 2013). 

One of most puzzling aspects of galaxy evolution is the tight
correlation between black hole mass and galaxy properties such as
bulge stellar mass and velocity dispersion (Magorrian et al. 1998;
Ferrarese \& Merritt 2000). This correlation suggests a physical 
connection between black hole accretion on sub-parsec (pc) scales and
galaxy-wide star formation and gas accretion on kiloparsec (kpc)
scales. The most likely explanation is quasar feedback, but the
details of how this operates as a function of cosmic time are still to
be determined (Cattaneo et al. 2009). Observationally, this topic can
be studied by measuring the global growth rates of black holes and
galaxies and the ratio of black hole to stellar mass in high redshift
galaxies. The black hole masses of quasars can be measured from the
dynamics of gas in the broad line region which is only $\sim 1$\,pc
from the black hole (Wandel 1999).

As well as studying the physical properties of star-forming gas in
high-redshift galaxies, millimeter interferometry can be used to probe
the gas dynamics. A particularly useful application is that of quasar
host galaxy dynamical masses to measure the ratio of black hole to
galaxy mass in the early universe (Walter et al. 2004). Observations
in the rest-frame UV or optical are hampered by the overwhelming
brightness of the quasar point-source (e.g. Mechtley et
al. 2012). Wang et al. (2010) showed that CO line widths of the most
optically luminous $z\approx6$ quasars indicate ratios of black hole
to galaxy mass a factor of 10 on average greater than found in the
local universe. This could signal that black holes grow much more
rapidly than their host galaxies within the first billion
years. However, there is a selection bias detailed in depth by Lauer
et al. (2007) which suggests that the most luminous quasars will be
biased to high black hole mass due to scatter in the correlation. In
order to check whether this bias affects the conclusions of Wang et
al. (2010) it is important to determine the black hole to galaxy mass
relationship for the more common, lower luminosity quasars at
$z\approx 6$ such as those identified in the Canada-France High-z
Quasar Survey (CFHQS; Willott et al. 2010a).

An alternative approach to studying the co-evolution of galaxies and
black holes is to determine the star formation rate of active galaxies
over cosmic time (Carilli et al. 2001; Omont et al. 2003; Priddey et
al. 2003; Wang et al. 2008, 2011a; Lutz et al. 2010; Serjeant et al. 2010;
Bonfield et al. 2011; Omont et al. 2013). These studies suggest
positive evolution in the star formation rate (SFR) at a fixed quasar
luminosity from $z=0$ to $z\approx 2$ and approximately constant or a
mild decline at higher redshift. However, in these studies, most
high-redshift quasars are undetected and conclusions have to be based
on statistical detections of stacked sub-samples. ALMA now provides an
opportunity to measure the SFR at least an order of magnitude fainter than
previous observations and revolutionize our understanding of the
co-evolution of black holes and their host galaxies.

In this paper we present an ALMA study of the interstellar medium of
the host galaxies of two $z=6.4$ CFHQS quasars. These quasars were
selected for study based on their high redshift (they are two of the
four highest redshift published quasars), faint absolute magnitude
($M_{1450} \geq -25$) and black hole masses of $\sim 10^{8}\,{\rm
  M}_\odot$. Section 2 details the new observations. The results are
presented in Section 3. Section 4 contains a discussion of the
results.  Cosmological parameters of $H_0=70~ {\rm
  km~s^{-1}~Mpc^{-1}}$, $\Omega_{\mathrm M}=0.27$ and
$\Omega_\Lambda=0.73$ (Komatsu et al. 2011) are assumed throughout.

\section{Observations}

CFHQS\,J021013-045620 (hereafter J0210-0456) and CFHQS\,J232908-030158
(hereafter J2329-0301) were observed with ALMA between June and August
2012 in {\it Early Science} project 2011.0.00243.S. The number of
12\,m diameter antennae in use ranged from 17 to 24 with a typical
longest baseline of 400\,m. Observations of the science targets were
interleaved with nearby phase calibrators, J0217+017 and
J2323-032. Uranus was used as the amplitude calibrator and 3C446 as
the bandpass calibrator. Total on-source integration times were
8000\,s for J0210-0456 and 8500\,s for J2329-0301.

\begin{figure}[t]
\includegraphics[angle=0,scale=0.30]{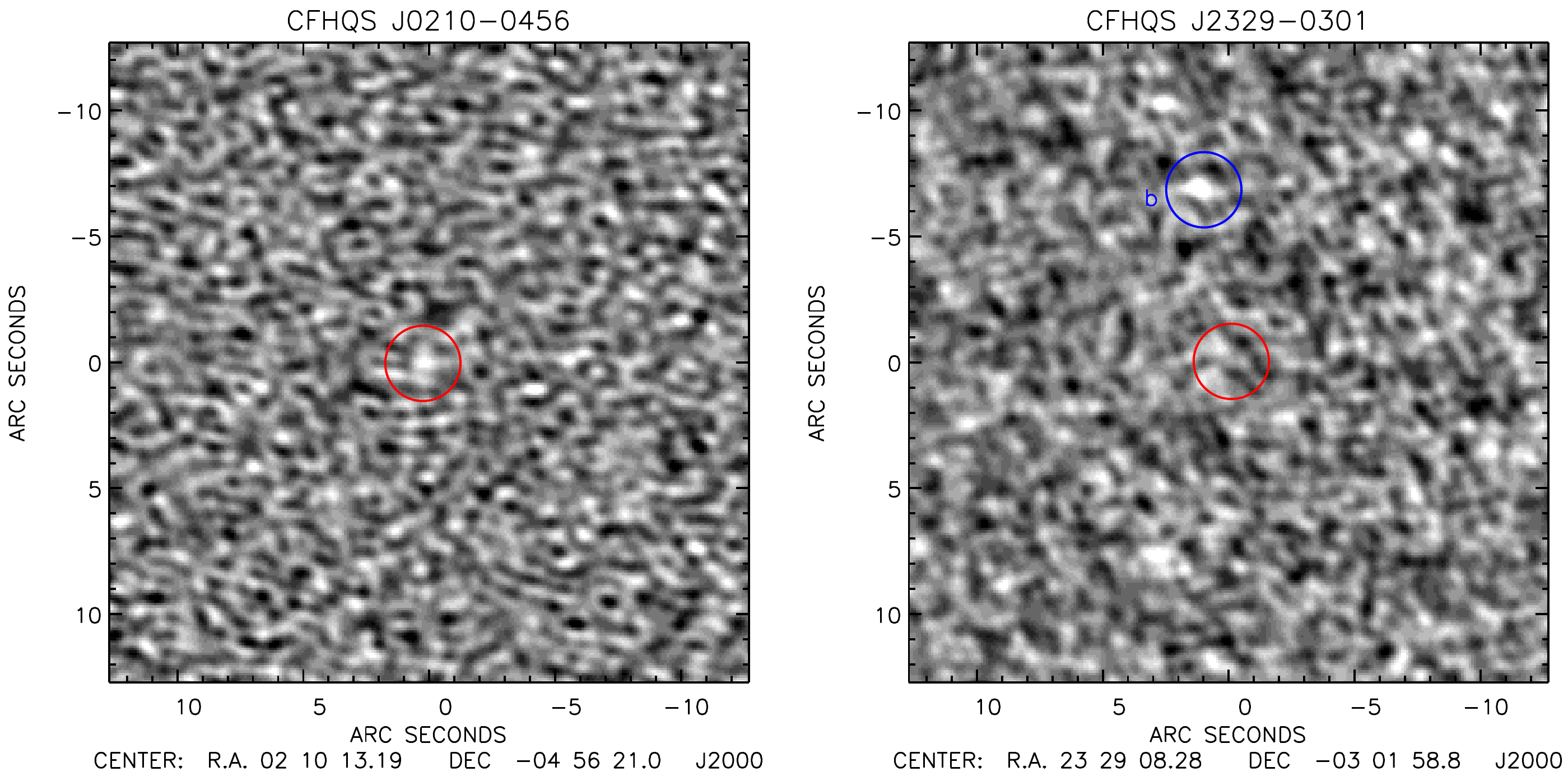}
\caption{ALMA 1.2\,mm continuum images generated from the three line-free basebands for each of the two quasar fields. The greyscale ranges from $-3\,\sigma$ (black) to $+3\,\sigma$ (white) where $\sigma=35,30 \, \mu$Jy\,beam$^{-1}$ for J0210-0456 (left) and  J2329-0301 (right), respectively. Red circles show the expected source positions (circles do not indicate the positional uncertainty). There is a $3.4\,\sigma$ detection for J0210-0456 and no detection for J2329-0301. The strongest source in the field of J2329-0301 is marked with a blue circle and labelled 'b'. It is co-incident with a blue galaxy in the optical images of Willott et al. (2007). }
\label{fig:continmaps}
\end{figure}
 
The band 6 (1.3\,mm) receivers were set up so that one of the four
basebands (each of width 1.875\,GHz) was centred on the expected
location of the redshifted \cii\ transition ($\nu_{\rm
  rest}$=1900.5369 GHz). The redshifts adopted were those of the
low-ionization broad \mgii\ lines of the quasars measured by Willott
et al. (2010b). Previous studies of high-redshift quasars have shown
relatively small offsets ($1\sigma$ dispersion 270\,km\,s$^{-1}$)
between \mgii\ and the systemic redshift (Richards et al. 2002). The
remaining three spectral windows were placed nearby to sample the 1.2\,mm
continuum. Each baseband is sampled by 120 channels of width
15.625\,MHz (equivalent to $\approx 18$\,km\,s$^{-1}$).

Data processing was performed by staff at the North American ALMA
Regional Center using the {\small CASA} software package\footnotemark.
The three line-free spectral windows were combined to generate 1.2\,mm
continuum images. Both the continuum maps and spectral line datacubes
were spatially sampled with $0\farcs1$ pixels. The synthesized beams
are $0\farcs77$ by $0\farcs52$ for J0210-0456 and $0\farcs73$ by
$0\farcs63$ for J2329-0301. The noise level reached in a 2 channel bin
(31.25\,MHz) is 0.22\,mJy\,beam$^{-1}$ for J0210-0456 and
0.23\,mJy\,beam$^{-1}$ for J2329-0301.

\footnotetext{http://casa.nrao.edu}

\section{Results}

\subsection{Far-infrared luminosity}
\label{lfir}

The 1.2\,mm continuum luminosity of $z=6.4$ sources probes rest-frame
$160\,\mu$m radiation, on the Rayleigh-Jeans tail side of the typical
star-forming galaxy dust spectral energy distribution (SED; Lagache et al. 2005). This
makes it an excellent proxy for the total far-infrared luminosity
($L_{\rm FIR}$; integrated luminosity over $42.5 - 122.5\,\mu$m) which
is a reliable tracer of the star formation rate due to dust heated by
young stars. In the most ultraviolet-luminous quasars (such as those
at $z\sim 6$ in the SDSS), there is often a substantial contribution
to $L_{\rm FIR}$ from dust heated by the AGN (Wang et al. 2008). The
CFHQS quasars are an order of magnitude less UV-luminous than SDSS
quasars and therefore should have a correspondingly lower contribution
from AGN-heated dust, allowing continuum observations to probe lower
star-formation rates.

%Now using pdflatex so .ps (or .eps) plots from IDL are opened in preview (so converted to .pdf), then use selection tool and crop to desired size
%For large plots (outside letter size) use ps2pdf -sPAPERSIZE=a3 image.ps 

\begin{figure*}[t]
\includegraphics[angle=0,scale=1.0]{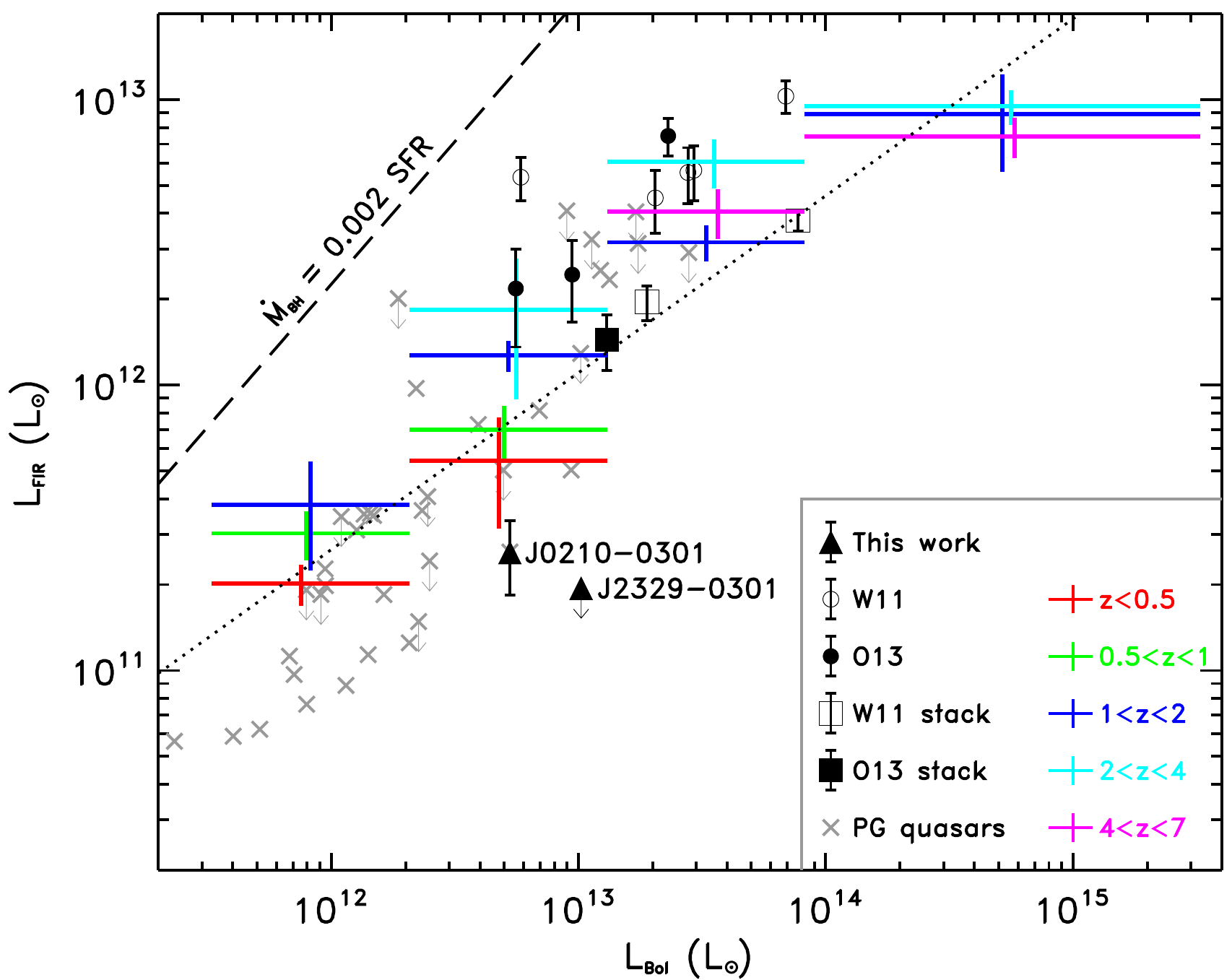}
\caption{Far-infrared luminosity versus AGN bolometric luminosity for $z\approx 6$ quasars. The two CFHQS quasars observed with ALMA in this paper are shown with triangles. Previous detections from Wang et al. (2011a), mostly of SDSS quasars with a few quasars from other surveys, are shown as open circles. Previous detections of three CFHQS quasars are shown as filled circles (Omont et al. 2013). The squares show stacked averages from Wang et al. (2011a) and Omont et al. (2013). A complete sample of local PG quasars are shown with gray crosses (Hao et al. 2005). Colored lines show stacked averages in bins in redshift and $L_{\rm Bol}$ for quasars from the H-ATLAS survey and other far-IR/mm data (Serjeant et al. 2010). The dotted line is a fit to high-redshift ($2<z<7$) stack averages (Wang et al. 2011a). The dashed line converts $L_{\rm Bol}$ to black hole accretion rate and $L_{\rm FIR}$ to star formation rate such that these grow in parallel to match the local $M_{\rm BH}/M_{\rm stellar}$ relationship (Tundo et al. 2007).}
\label{fig:lbollfir}
\end{figure*}

The ALMA 1.2\,mm continuum images generated from the three spectral
windows that did not include the \cii\ line were analyzed to determine
their flux-densities. These images are shown in Figure
\ref{fig:continmaps} where the expected locations of the quasars are
identified by red circles. J0210-0456 is detected at $3.4\,\sigma$
with $f_{\rm 1.2mm} = 120 \pm 35 \,\mu$Jy. At this significance level
it is not possible to determine whether the source is spatially
resolved. J2329-0301 is undetected with no hint of positive flux at
the quasar location. The most significant continuum source in the
field ($7''$ north of the quasar, labeled 'b') is identified as a blue galaxy at
much lower redshift in the optical imaging of Willott et al. (2007).

This continuum flux-density was converted to a far-infrared luminosity
assuming a typical SED for high-redshift star-forming galaxies. To
make meaningful comparison with previous results (in particular Wang
et al. 2008; 2011a, Omont et al. 2013) we adopt a greybody spectrum
with dust temperature, $T_{\rm d} =47$\,K and emissivity index,
$\beta=1.6$. We note that our faint sources have much lower mm fluxes
than the typical sources used to determine these parameters. If our
sources instead have dust temperature closer to that of nearby
luminous infrared galaxies (LIRGs, $10^{11} - 10^{12} \, {\rm
  L}_\odot, T_{\rm d}\approx 33$\,K, U et al. 2012) then the values of
$L_{\rm FIR}$ would be {$3\times$ lower}. For the remainder of this
paper, uncertainties on $L_{\rm FIR}$ (and inferred SFR) only include
the flux measurement uncertainties, not that of the dust temperature.

The far-IR luminosity of J0210-0456 is $(2.60 \pm 0.76) \times 10^{11}
\, {\rm L}_\odot$. J2329-0301 is undetected with $L_{\rm FIR} <1.9 \times
10^{11} \, {\rm L}_\odot $ ($3\,\sigma$ limit). We note the incredible
sensitivity of these early ALMA observations that reach the lower
end of the LIRG classification in the early universe at $z=6.4$.

We now consider the relationship between $L_{\rm FIR}$ and the quasar
bolometric luminosity $L_{\rm Bol}$ for high-redshift quasars. $L_{\rm
  Bol}$ in this case is for the quasar component of the galaxy and
assumes a typical bolometric correction from the rest-frame UV
luminosity at 1450 \AA\ of a factor of 4.4 (Richards et al. 2006). $L_{\rm Bol}$
does not include any excess FIR luminosity above that of the typical
quasar. It is still a matter of debate as to how much of the typical
quasar far-IR emission is due to dust heated by the AGN, compared to
dust heated by a starburst (Haas et al. 2003; Hao et al. 2005; Netzer et al. 2007; Lutz
et al. 2010).

Because most high-redshift quasars have not been detected in mm
continuum with the sensitivity level of previous studies, the
relationship between $L_{\rm FIR}$ and $L_{\rm Bol}$ has been based on
stacking of sub-samples with different bolometric luminosity
ranges. Wang et al. (2011a) showed that the stacks based on several
samples at $2< z <7$ could be fit by the relationship $L_{\rm FIR}
\propto L_{\rm Bol}^{0.6}$.  Omont et al. (2013) found that the
stacked average from 1.2\,mm MAMBO observations of CFHQS $z\approx 6$
quasars also lie on this relationship. The non-linear nature and
significant scatter (for those detected so far) is interpreted in an
evolutionary scenario where both star formation rate and black hole
accretion are dependent upon dark matter halo mass, but with a lack of
synchronization in the rates of these processes. The galaxies with the
lowest $L_{\rm FIR}$ at a given $L_{\rm Bol}$ are expected to have a
significant fraction of their $L_{\rm FIR}$ due to quasar-heated dust
(Netzer et al. 2007).

Figure \ref{fig:lbollfir} shows previously published data for
individually-detected $z\approx 6$ quasars and stacked averages from
Wang et al. (2011a) and Omont et al (2013). The dotted line is the
relationship between $L_{\rm FIR}$ and $L_{\rm Bol}$ found by Wang et
al. (2011a) for stack averages of quasars at $2< z <7$. Note that both
the stacked averages and relationship from Wang et al. 2011a have been
renormalized according to the bolometric correction adopted here (see
Omont et al. 2013 for more details). Gray crosses are a complete
sample of low-redshift ($z<0.5$) optically-selected Palomar-Green (PG)
quasars (Hao et al. 2005). $L_{\rm FIR}$ for PG quasars has been
estimated as $2 \,\times$ the luminosity at $60\,\mu$m (Lawrence et
al. 1989). Note that many of the highest luminosity (most distant)
quasars in the PG sample are undetected at $60\,\mu$m and only have
upper limits on $L_{\rm FIR}$. As noted by Wang et al. (2011a), the
$z\approx 6$ stacked averages lie close to the correlation exhibited
by low-redshift quasars, indicating no enhancement in SFR at high
redshift for a given quasar luminosity.

Serjeant et al. (2010) used {\it Herschel} imaging of quasars in the
H-ATLAS survey plus supplementary published IR and mm data to
determine the average quasar far-infrared luminosity as a function of
both redshift and quasar luminosity. Their data (for all bins
containing 10 or more quasars) is shown in Figure \ref{fig:lbollfir}
using an $I$ band bolometric correction of 12.0 (Richards et al. 2006)
and $L_{\rm FIR} = 1.75\,\times$ the luminosity at $100\,\mu$m. These
data show a similar correlation of the two luminosities as found for
the PG and $z\approx 6$ samples. However, Serjeant et al. do find a
positive correlation between $L_{\rm FIR}$ and redshift up to
$z\approx 3$ that does not continue up to the $z=6$ data of Wang et
al. (2011a) and Omont et al. (2013). A positive correlation between
$L_{\rm FIR}$ and redshift up to $z\approx 2$ was also observed by
Bonfield et al. (2011).

The values for the two CFHQS quasars with ALMA data are also plotted
on Figure \ref{fig:lbollfir}. The detection of J0210-0456 shows its
$L_{\rm FIR}$ to be a factor of $\approx 3$ lower than the stacked
average relationship and below the stacked averages from H-ATLAS at
all redshifts. The non-detection of J2329-0301 corresponds to $L_{\rm
  FIR}$ at least a factor of $10$ lower than the stacked average from
the full CFHQS sample (Omont et al. 2013) and substantially below the
H-ATLAS averages.

The far-infrared luminosity can be used to derive the SFR assuming the
relation in Kennicutt (1998) with a Salpeter (1955) initial mass function (IMF). For J0210-0456, SFR\,$=48\, {\rm
  M}_\odot\,{\rm yr}^{-1}$ and for J2329-0301, SFR\,$<40\, {\rm
  M}_\odot\,{\rm yr}^{-1}$ ($3\,\sigma$ limit). For both these
objects, the assumption in deriving SFR is that there is no
contribution to $L_{\rm FIR}$ from quasar-heated dust. These quasars
lie close to the lower range of $L_{\rm FIR}$ where it
has been suggested that the majority of the cool dust is heated by the
quasar (Netzer et al. 2007). If this is the case then the SFR would be
even lower. The very low SFR implied for the host galaxy of J2329-0301
is surprising given that it has a $2.5 \times 10^{8}\, {\rm M}_\odot$
black hole accreting at the Eddington rate (Willott et al. 2010b) and
a very luminous, spatially-extended \lya\ halo (Goto et al. 2009;
Willott et al. 2011).

\begin{figure}[t]
\includegraphics[angle=0,scale=0.48]{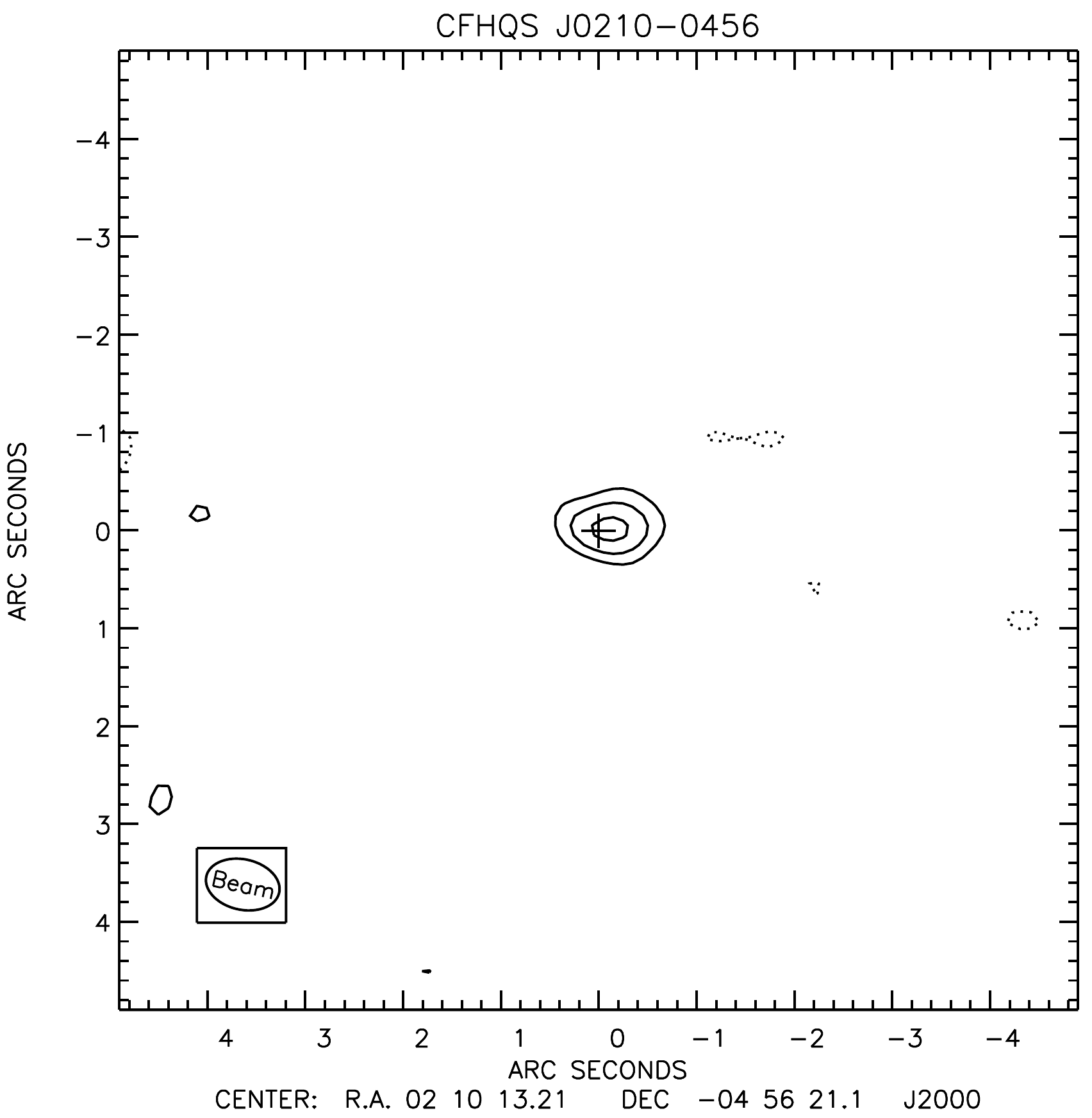}
\caption{Continuum-subtracted \cii\ line emission for CFHQS\,J0210-0456 integrated over 15 channels containing the \cii\ line. Contours are at $[-3,3,5,7]\times \sigma$ where $\sigma=0.03$\,Jy\,km\,s$^{-1}$\,beam$^{-1}$.}
\label{fig:linemap}
\end{figure}

\subsection{\cii\ luminosity}

The datacubes of the spectral windows containing the expected \cii\ emission lines for the two quasars were inspected for line emission. A line was easily detected for J0210-0456, but not for J2329-0301. Figure \ref{fig:linemap} shows the image of \cii\ emission for J0210-0456 integrated over the 15 channels (each of width 15.625\,MHz) that show line emission. Continuum emission has been subtracted off using the continuum image of Figure \ref{fig:continmaps}. The source is elongated east-west, although note this is close to the major axis of the elongated beam. The spatial structure will be discussed further in the following section. No other \cii\ emitters at the same redshift are seen in the field.

\begin{figure}[t]
\includegraphics[angle=0,scale=0.47]{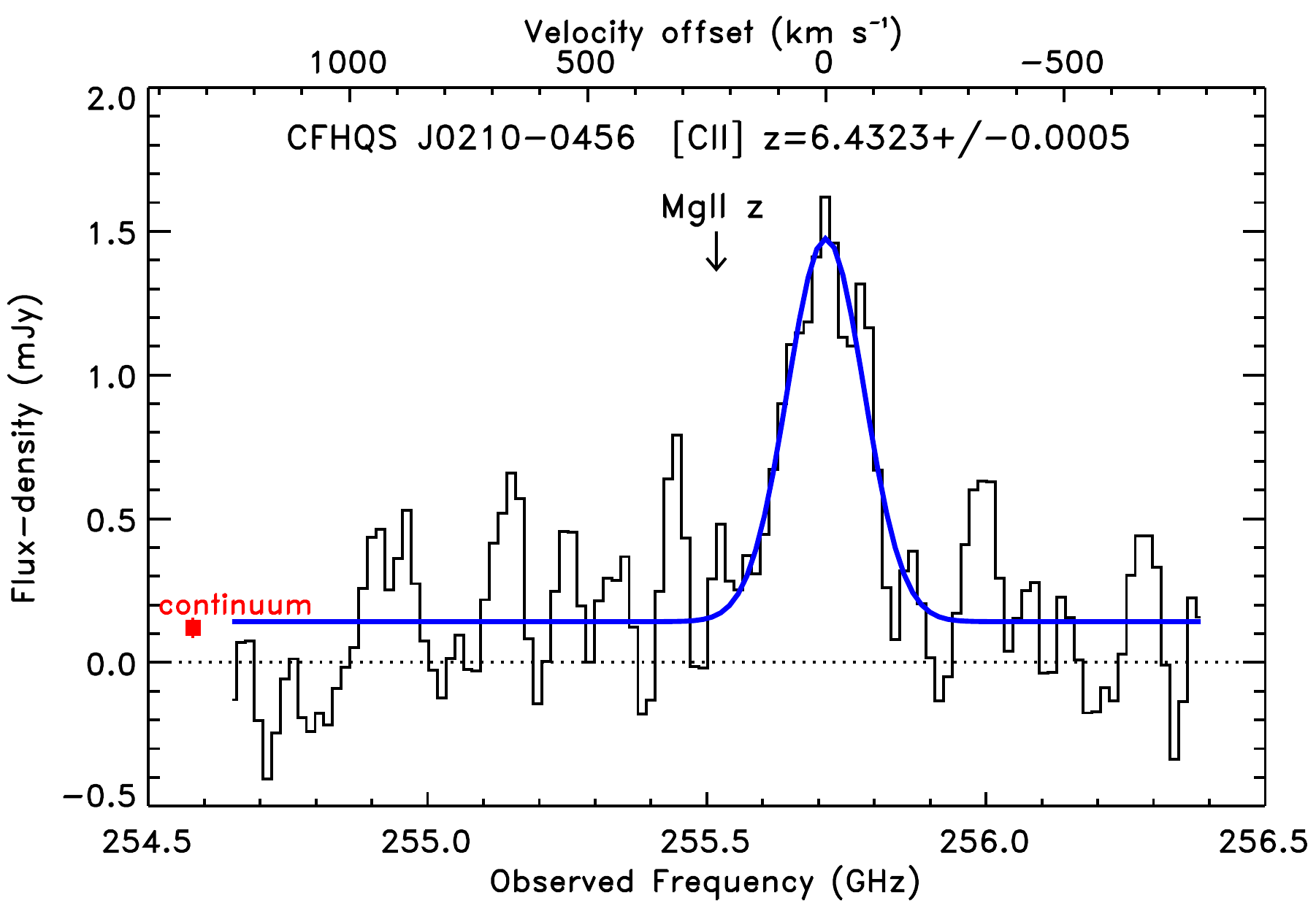}
\caption{\cii\ spectrum for CFHQS\,J0210-0456 overlaid with best fit Gaussian plus continuum model (blue). The red square with error bar is the continuum flux measured from the three line-free basebands. The arrow marked \mgii\ shows the redshift measured from the quasar broad line region. }
\label{fig:linespec0210}
\end{figure}

The spectrum of J0210-0456 is plotted in Figure
\ref{fig:linespec0210}. The \cii\ line is offset from the broad
ultraviolet \mgii\ emission line by $230$\,km\,s$^{-1}$. Note that the
rms observational uncertainty on the \mgii\ redshift is
$160$\,km\,s$^{-1}$, so the redshifts of \cii\ and \mgii\ are
consistent. We take the redshift of $z_{\rm [CII]} = 6.4323 \pm
0.0005$ to be the systemic redshift because it is measured to much
higher accuracy than \mgii\ and is associated with star formation in
the host galaxy rather than gas in the circum-quasar environment. A
simple Gaussian plus constant fit was made to the observed spectrum. A
Gaussian with FWHM = $189 \pm 18$\,km\,s$^{-1}$ provides a good fit to
the line profile. The best-fit constant is positive showing a non-zero
continuum level that is consistent with the continuum flux-density of $120\,\mu$Jy
measured from the three line-free basebands in Section \ref{lfir}.

The improved systemic redshift for J0210-0456 allows an improvement in
the determination of the size of the highly-ionized near-zone. The
size of the region ionized by the quasar depends upon several factors
including the neutral hydrogen fraction of the intergalactic medium
when the quasar first turned on (Madau \& Rees 2000; Cen \& Haiman
2000) and therefore can be used to probe cosmic reionization. Willott
et al. (2010b) used the \mgii\ redshift of this quasar to determine a
near-zone size of 1.7 proper Mpc, which is lower than any other
$z\approx 6$ quasar except for lineless or broad absorption line
quasars (Fan et al. 2006; Carilli et al. 2010). Using the new redshift of $z_{\rm
  [CII]} = 6.4323$ gives a near-zone size of 1.4 proper Mpc, even
lower than previously calculated. Correcting for the known
luminosity-dependence of $R\propto L^{1/3}$ makes the size only
slightly lower than the typical size for more luminous $z>6.1$ quasars
of 5\,Mpc (Carilli et al. 2010).

\begin{figure}[t]
\includegraphics[angle=0,scale=0.47]{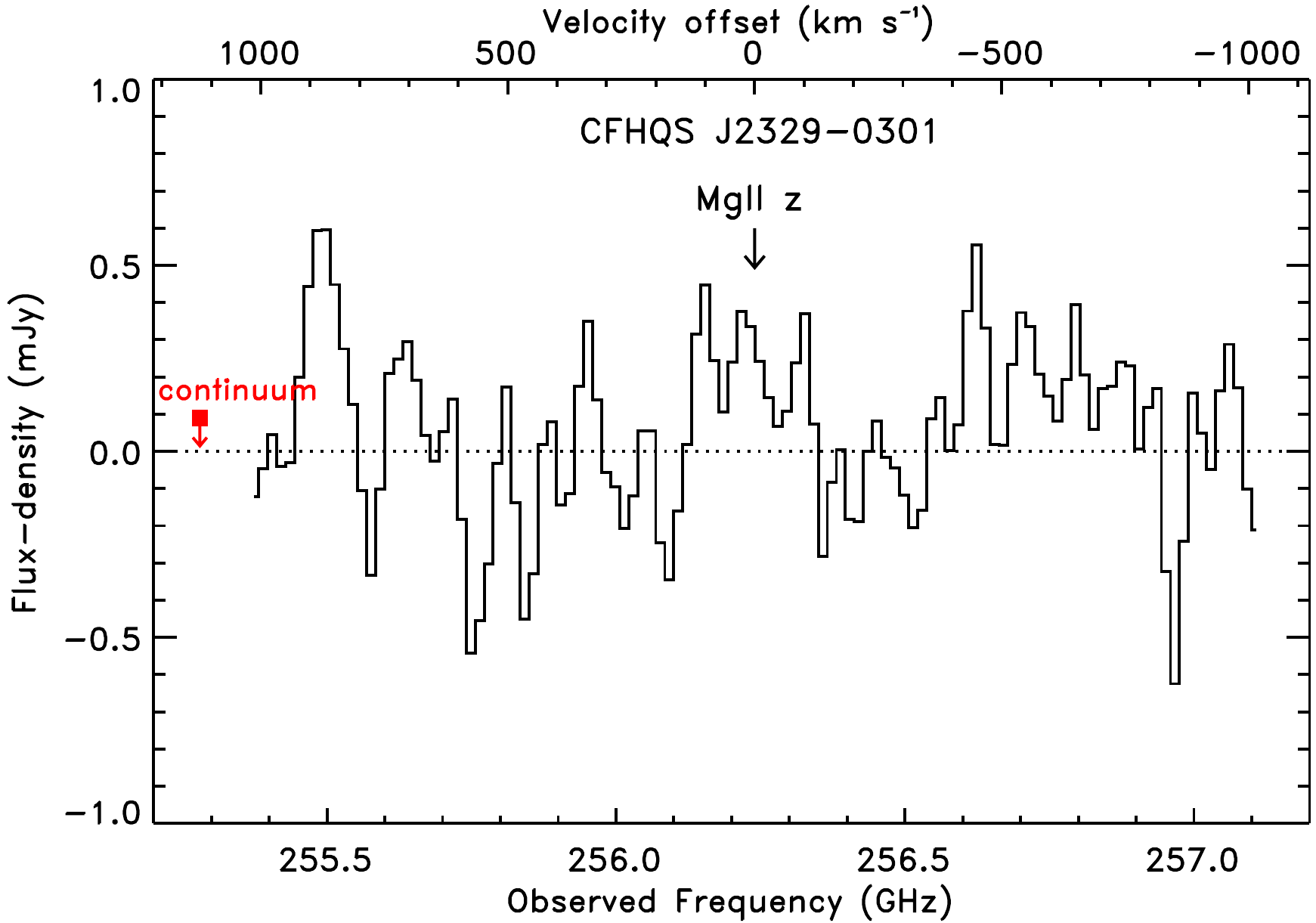}
\caption{\cii\ spectrum for CFHQS\, J2329-0301. There is no convincing detection of the line for this quasar. The red square with downward arrow is the limit on the continuum flux from the three line-free basebands. }
\label{fig:linespec2329}
\end{figure}

The spectrum of J2329-0301 is plotted in Figure
\ref{fig:linespec2329}. It can be seen that there is not strong
evidence for a measurable \cii\ emission line. There is weak positive
flux at the \mgii\ redshift that may correspond to real emission, but
it is very uncertain so we assume here a non-detection.

The \cii\ line flux of J0210-0456 was determined by integrating over
the channels containing the line and subtracting off the continuum
component. This line flux was then converted to a line luminosity at
the measured redshift. For J2329-0301 a $3\sigma$ upper limit for the
\cii\ flux and luminosity was determined by assuming a spatially
unresolved Gaussian with FWHM = $300$\,km\,s$^{-1}$. This line width
is somewhat broader than that observed for J0210-0456 but narrower
than CO line widths for SDSS $z \approx 6$ quasars (Wang et
al. 2010). Measurements of emission line and continuum parameters
derived from these data are quoted in Table \ref{tab:data}.

\begin{table}
\begin{center}
\caption{Millimeter data for CFHQS $z\approx 6.4$ quasars\label{tab:data}}
\begin{tabular}{lll}
\tableline\tableline
& CFHQS\,J0210-0456 & CFHQS\,J2329-0301\\
\tableline
%\mgii\ redshift 
$M_{\rm BH}$ &  $(8.0^{+5.5}_{-4.0})\times 10^{7} M_\odot \,^{\rm a}$ &  $(2.5^{+0.4}_{-0.4})\times 10^{8} M_\odot \,^{\rm a}$\\
$z_{\rm MgII}$ &  $6.438 \pm 0.004 ~^{\rm a}$ &  $6.417 \pm 0.002 \,^{\rm a}$ \\
%\cii\ redshift 
$z_{\rm [CII]}$ &  $6.4323 \pm 0.0005$ & $-$\\
FWHM$_{\rm [CII]}$ & $189 \pm 18$\,km\,s$^{-1}$ & $-$\\
%\cii\ line flux 
$I _{\rm [CII]}$  & $0.269  \pm 0.037$\,Jy\,km\,s$^{-1}$ & $<0.10$\,Jy\,km\,s$^{-1} \,^{\rm b}$\\
%\cii\  luminosity 
$L_{\rm [CII]}$ & $(3.01 \pm 0.41) \times 10^8  \,{\rm L}_\odot$ & $<1.1 \times 10^8  \,{\rm L}_\odot \,^{\rm b}$ \\
%CO line flux 
$I _{\rm CO(2-1)}$  & $<0.014$\,Jy\,km\,s$^{-1} ~^{\rm c}$ & $-$\\
%CO luminosity 
$L_{\rm CO(2-1)}$ & $<1.6 \times 10^7  \,{\rm L}_\odot ~^{\rm c}$ & $-$ \\
%mm flux-density 
$f_{\rm 1.2mm}$ & $120 \pm 35 \,\mu$Jy & $<90\, \mu$Jy \,$^{\rm d}$\\
%Far-IR luminosity 
$L_{\rm FIR}$  & $(2.60 \pm 0.76) \times 10^{11} \, {\rm L}_\odot$  & $<1.9 \times 10^{11} \, {\rm L}_\odot  \,^{\rm d}$\\
SFR & $48 \pm 14\, {\rm M}_\odot\,{\rm yr}^{-1}$  & $<40\, {\rm M}_\odot\,{\rm yr}^{-1}$ \,$^{\rm d}$\\
$L_{\rm [CII]} / L_{\rm FIR}$ & $(1.15 \pm 0.32) \times 10^{-3}$  & $-$ \\ 
$L_{\rm CO(1-0)} / L_{\rm FIR}$ & $<8.5 \times 10^{-6}$  & $-$ \\ 
\tableline
\end{tabular}
\end{center}
{\sc Notes.}---\\
$^{\rm a}$ Derived from \mgii\ $\lambda2799$ observations (Willott et al. 2010b).\\
$^{\rm b}$ 3$\sigma$ upper limit assuming spatially unresolved and line width FWHM=300\,km\,s$^{-1}$.\\
$^{\rm c}$ 3$\sigma$ upper limit from observations in Wang et al. (2011b) assuming spatially unresolved and FWZI=300\,km\,s$^{-1}$.\\
$^{\rm d}$ 3$\sigma$ upper limit assuming spatially unresolved.
\end{table}

The [CII] line is primarily produced in photo-dissociation regions and
is strongly dependent on the interstellar radiation field (Stacey et
al. 1991). The ratio $L_{\rm [CII]} / L_{\rm FIR}$ has an inverse
correlation on the radiation field strength and has been widely
studied at lower redshift. It has been found that the ratio has an
inverse correlation with $L_{\rm FIR}$ (Luhman et
al. 2003). Graci\'a-Carpio et al. (2011) found this inverse correlation
is even tighter if one normalises the far-IR luminosity by the
molecular gas mass $M_{H_{2}}$. They attributed this to $L_{\rm
FIR}/M_{H_{2}}$ being more closely related to the physical properties
of the clouds such as density and temperature.

\begin{figure}
\includegraphics[angle=0,scale=0.465]{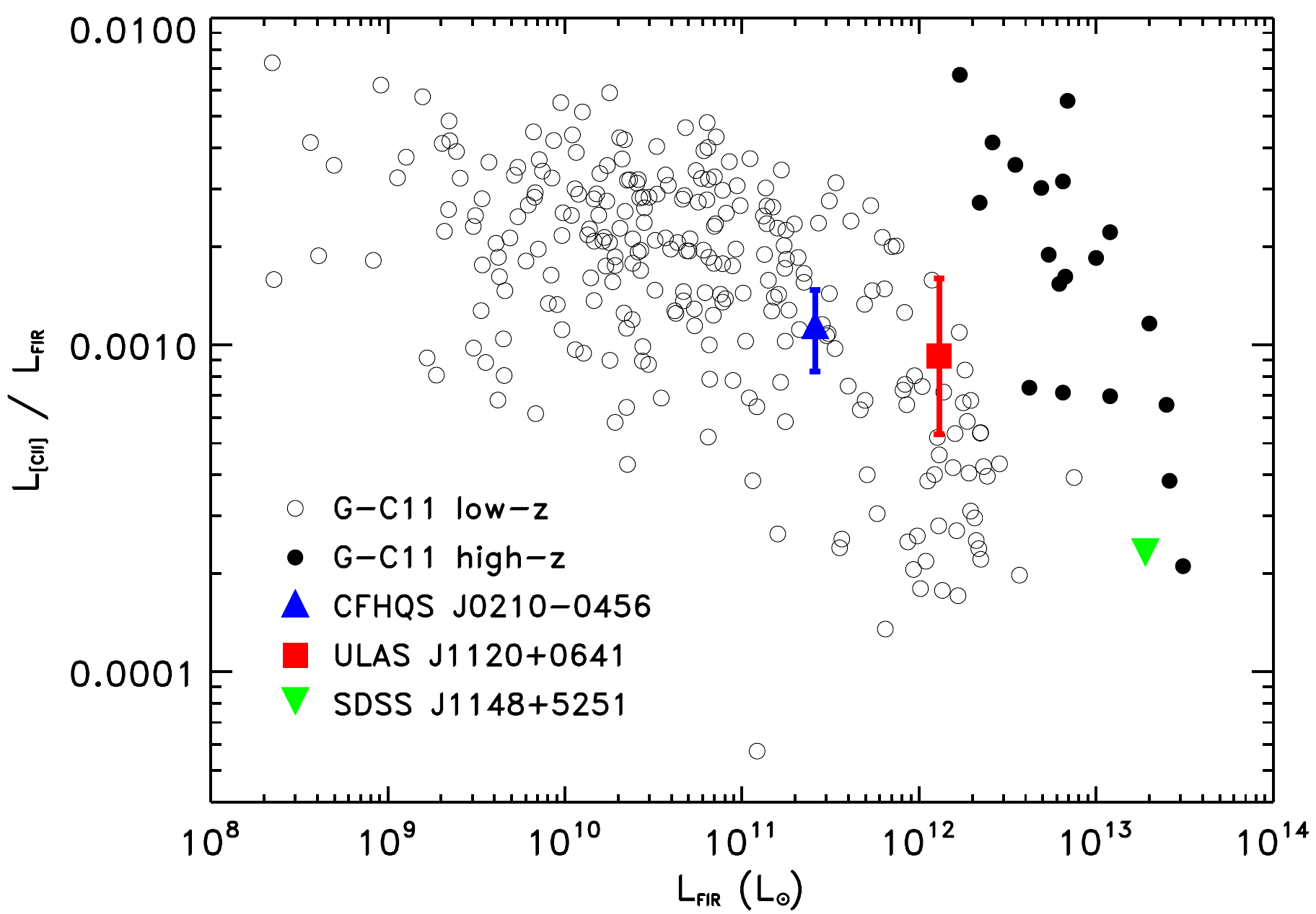}
\caption{Ratio of \cii\ to far-infrared luminosity versus far-infrared luminosity. Open circles show low-redshift ($z<0.4$) and filled circles high-redshift ($1<z<5$) galaxies from the collection by Garci\'a-Carpio et al. (2011 and in prep.). Detections of quasars at $z>6$ are shown individually: J0210-0456 (this paper), SDSS\,J1148+5251 (Maoilino et al. 2005) and ULAS\,J1120+0641 (Venemans et al. 2012). Note the outstanding high-redshift $L_{\rm FIR}$ sensitivity of ALMA in this {\it Early Science} observation with a modest integration time. }
\label{fig:lfirlcii}
\end{figure}

Figure \ref{fig:lfirlcii} shows $L_{\rm [CII]} / L_{\rm FIR}$ against
$L_{\rm FIR}$ for a compilation of low ($z<0.4$) and high ($1<z<5$)
redshift galaxies (Graci\'a-Carpio et al. 2011 and in prep.). The
horizontal offset between the low and high redshift sources was
attributed by Graci\'a-Carpio et al. (2011) to the higher molecular
gas mass (for a given $L_{\rm FIR}$) at high redshift (e.g. Tacconi et
al. 2010). Also plotted on Figure \ref{fig:lfirlcii} are data for
$z>6$ quasars. SDSS\,J1148+5251 has a very high $L_{\rm FIR}$ and
falls along the sequence of $2<z<5$ high-redshift galaxies (Maiolino
et al. 2005). ULAS\,J1120+0641 (Venemans et al. 2012) and
CFHQS\,J0210-0456 have more moderate $L_{\rm FIR}$ and fall within the
region occupied by low-redshift galaxies. In the interpretation of the
offset being due to higher molecular gas mass at high-redshift, this
would suggest that not all high-redshift quasars exist in star-forming
hosts with higher gas masses than at low redshift. The horizontal
offset that is so striking in Figure \ref{fig:lfirlcii} is at least
partially due to a selection effect where previous facilities did not
have the sensitivity to detect more moderate $L_{\rm FIR}$ at high
redshifts and only ultraluminous continuum sources were followed up
with \cii\ observations. It is likely there is a large population of
hitherto undetected high-redshift galaxies with properties like
J0210-0456.

Wang et al. (2011b) reported Very Large Array observations aimed at
detecting the CO ($2-1$) emission from J0210-0456. The object was not
detected and a line flux upper limit assuming a full-width
zero-intensity of 800 \,km\,s$^{-1}$ was reported. We have
recalculated the line flux limit for the same width as the observed
\cii\ line (FWZI = $300$\,km\,s$^{-1}$).  The $3\sigma$ upper limit on
the line flux is then $<0.014$\,Jy\,km\,s$^{-1}$ giving a CO ($2-1$)
line luminosity limit of $L_{\rm CO (2-1)}< 1.6 \times 10^7 \,{\rm
  L}_\odot$. To compare with other works that usually quote the
ground-state CO transition we assume a luminosity ratio of CO ($2-1$) / CO ($1-0)
= 7.2$ (Stacey et al. 2010; Papadopoulos et al. 2012). Therefore the CO
($1-0$) limit for J0210-0456 is $L_{\rm CO (1-0)} < 2.2 \times 10^6
\,{\rm L}_\odot$ and the ratio $L_{\rm CO (1-0)} / L_{\rm FIR} <8.5
\times 10^{-6}$. The limit on this ratio is an order of magnitude
higher than the values for typical ultraluminous high-redshift
galaxies and AGN (De Breuck et al. 2011) showing that much deeper
observations are required to detect the molecular gas in galaxies such
as these, highlighting how \cii\ observations with ALMA are the best
way to probe the obscured interstellar medium in typical high-redshift
galaxies.

\subsection{\cii\ dynamics and spatial extent}

Wang et al. (2010) showed that CO line widths of $z\approx6$ SDSS
quasars indicate ratios of black hole to galaxy mass typically a
factor of 10 greater than in the local universe. This fits with the
results presented in Section \ref{lfir} where it was found that
$z\approx6$ quasar host galaxies are growing their black holes at a
rate about $10\times$ faster than their stellar mass compared to the
local ratio. In this work we have measured the \cii\ line width for just one
$z=6.4$ quasar, so will only briefly discuss the ratio of black hole
to dynamical mass at $z\approx 6$ and defer a fuller investigation
until more $\sim 10^8 \,{\rm M}_\odot$ black hole host galaxies have
suitable mm interferometry data. 

CFHQS\,J0210-0456 has a \cii\ line FWHM of $189 \pm 18$\,km\,s$^{-1}$,
equivalent to $\sigma=80 \pm 8$\,km\,s$^{-1}$ for a Gaussian and
ignoring any inclination correction. Based on the local relationship
(Gultekin et al. 2009) a galaxy with $\sigma=80$\,km\,s$^{-1}$ would
be expected to have $M_{\rm BH} \approx 3 \times 10^6 \,{\rm
  M}_\odot$, a factor of 25 lower than the measured $M_{\rm BH} =
8\times 10^7 \,{\rm M}_\odot$. This difference is comparable to the
factor of $10$ found by Wang et al. for more luminous ($M_{\rm BH}
\sim 10^9 \,{\rm M}_\odot$) quasars. Even after taking account of
possible inclination effects (Carilli \& Wang 2006) this shows the
black holes in $z\approx 6$ quasars to be considerably more massive
than expected from the local relationship between black hole and
galaxy mass.

\begin{figure}
\includegraphics[angle=0,scale=0.255]{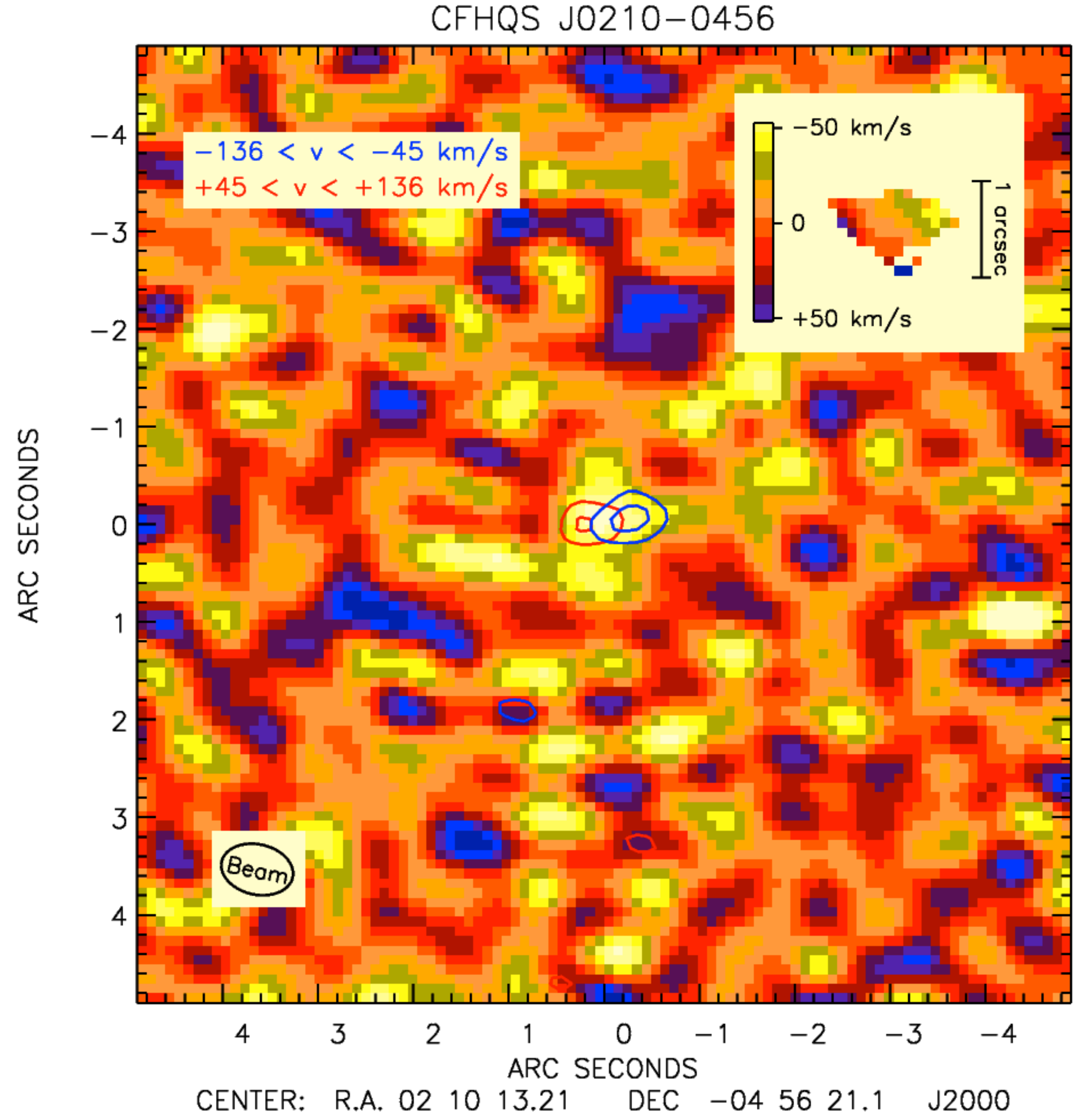}
\caption{The background shows the continuum map of J0210-0456 and ranges from $-3 \sigma$ (purple) to $+3\sigma$ (yellow). The blue and red contours near the central dust continuum source show \cii\ line emission maps for the blue ($-136 < v < -45$\,km\,s$^{-1}$) and red ($+45 < v < +136$\,km\,s$^{-1}$) wings, respectively. There is an offset of $0\farcs5$  between these peaks, either side of the continuum centroid. The inset panel shows a \cii\ peak velocity map for those pixels with sufficient flux to enable a Gaussian to be fitted. This map reveals a smooth velocity gradient across the source.}
\label{fig:redblu}
\end{figure}

To fully determine the gas kinematics requires spatially resolving the
line emission. This should be possible with the full ALMA array that
will offer spatial resolution reaching 20 milliarcseconds (mas), which is
equivalent to $120$\,pc. Walter et al. (2009) found that the \cii\
emission in SDSS\,J1148+5251 is concentrated within a radius of only
$750$\,pc of the nucleus. This is consistent with the small sizes of
luminous $z\approx 6.5$ Lyman Break galaxies that have effective radii
$\approx 800$\,pc (Ono et al. 2013). 

The \cii\ line image of J0210-0456 shown in Figure \ref{fig:linemap}
shows elongation along an E-W direction, roughly aligned with the
direction of the beam. The measured size of the source is $879 \pm
55$\,mas $\times$ $642 \pm 75$\,mas compared to a beam size of
$770$\,mas $\times$ $520$\,mas. We used the {\sc CASA IMFIT} task to
fit a deconvolved model image to the data. This results in a
deconvolved source of $521 \pm 248$\,mas $\times$ $224 \pm 297$\,mas
oriented at a position angle of 128 degrees east of north. The
intrinsic source size is not well constrained as it is only marginally
resolved in this image comprised of all fifteen spectral channels of
the \cii\ line.

A different story emerges when one considers the blue and red sides of
the \cii\ line separately. Maps were made using only the red and blue
wings (5 channels each) and excluding the centre of the line. Figure
\ref{fig:redblu} shows the continuum dust emission as the background
image. Superimposed on this are separate contours for the blue and red
sides of the \cii\ line. It can be seen that there is a spatial offset
of $0\farcs5$ (3\,kpc) between these peaks along a position angle
similar to the 128 deg major axis of the marginally resolved full
channel image. The inset panel of Figure \ref{fig:redblu} shows a
velocity-centroid map of the \cii\ line for pixels with sufficient
flux to enable a Gaussian emission line to be fitted. There is a clear
velocity gradient across the source along this same axis with
magnitude $\approx 100$\,km\,s$^{-1}$ across a size scale of $1\farcs0$
(6\,kpc). Whether this is due to rotation of a galaxy wide disk or has a
more complex origin in merging multiple components will require higher
spatial resolution observations. This is similar to the \cii\ velocity
gradients over this scale observed by Wang et al. (2013) in some
$z\sim 6$ SDSS quasars and quite different to the compact, intense,
central starburst observed in SDSS\,J1148+5251 (Walter et al. 2009).

\section{Discussion}

These observations with ALMA break new ground in their sensitivity to
moderately star-forming galaxies at high-redshift. Even with this
sensitivity, only one of the two $z\approx 6$ quasars was detected in
line and continuum emission. The far-IR luminosity of J0210-0456 is
$(2.60 \pm 0.76) \times 10^{11} \, {\rm L}_\odot$, whereas J2329-0301
remains undetected with $L_{\rm FIR} <1.9 \times 10^{11} \, {\rm
  L}_\odot $, significantly below the typical far-IR luminosity for a
quasar of this bolometric luminosity at any redshift. These low far-IR
luminosities are surprising and place strong constraints on the star
formation rates in the host galaxies of these Eddington-limited
quasars.

In the simplest black hole/galaxy co-evolution scenario, cosmic
stellar mass and black hole mass increase in lockstep, ending up at
the ratio observed in the local universe of $M_{\rm BH}/M_{\rm
  stellar}= 0.002$ (Tundo et al. 2007). Detailed simulations show that
in individual galaxies the phases of star formation and black hole
accretion are not synchronized (Li et al. 2007) and this accounts for
the significant scatter of points in Figure \ref{fig:lbollfir}. It is
trivial to calculate the relationship between star formation rate
(linearly related to $L_{\rm FIR}$) and black hole accretion rate
(linearly related to $L_{\rm Bol}$ and assuming an accretion
efficiency of 10\%) necessary to achieve $M_{\rm BH}/M_{\rm stellar}=
0.002$. This curve is plotted as the dashed line in the upper-left of
Figure \ref{fig:lbollfir}. All the optically-selected quasars observed
at mm wavelengths are growing their black holes at a relatively faster
pace than their stellar mass and this is not too surprising given that
they were selected by their quasar emission.  Galaxies should lie on
both sides of the dashed line during their lifetimes in order to reach
the local ratio at $z=0$. At $z\approx2$, mm-selected galaxies mostly
lie to the upper left of the line showing that they are growing their
stellar mass more rapidly than their black holes (Alexander et
al. 2005). Lutz et al. (2010) also showed that low-luminosity AGN from
deep X-ray surveys are found on the left side of such a line.

J2329-0301 is found to be growing its black hole at a rate of
$>100\times$ faster than its stellar mass in order to reach the local
ratio. It has a black hole accretion rate of $\dot M_{\rm BH}\,\approx
7\, {\rm M}_\odot\,{\rm yr}^{-1}$ and SFR\,$<40\, {\rm M}_\odot\,{\rm
  yr}^{-1}$ ($3\,\sigma$ limit, assuming $T_{\rm d}<47$\,K, no
AGN-heated cool dust and Salpeter IMF).  Khandai et al. (2012) report the results of
hydrodynamic simulations of $z\sim6$ quasar host galaxies. These show
SFR that range from $100$ to $1000 \, {\rm M}_\odot\,{\rm yr}^{-1}$ as
they evolve during the main black hole accretion phase. The
simulations are designed to match the properties of the most luminous
quasars from the SDSS. J2329-0301 has a black hole accretion rate
$\sim 3 \times$ lower than typical SDSS quasars and therefore scaling
down the lowest simulated SFR by this amount would result in
approximately the SFR limit observed for J2329-0301. This suggests
that J2329-0301 is observed at a rare phase where it has very low SFR
compared to its black hole accretion rate. In the Khandai et
al. (2012) simulations, the SFR usually drops at the epoch of peak
quasar accretion due to feedback heating the host galaxy
gas. J2329-0301 appears to have very effectively shut off star
formation. Although the far-IR data of high luminosity PG quasars has
many non-detections and the H-ATLAS data are just stacked averages
(Figure \ref{fig:lbollfir}), it would appear that few low redshift
quasars have such a low ratio of SFR to black hole accretion as
J2329-0301. This quasar is known to be surrounded by a luminous \lya\
halo at least 15\,kpc across (Goto et al. 2009; Willott et al. 2011),
which signifies a huge reservoir of diffuse gas likely photo-ionized by the
quasar. When this gas cools and accretes on to the galaxy, a further
bout of star formation is likely. Hayes et al. (2013) noted that the \lya\ emission from galaxies with low dust content tends to be more extended than that in dusty galaxies, but do not provide a simple explanation for why this happens.  J2329-0301 certainly fits this pattern with a low dust content as measured by thermal dust emission and a very extended  \lya\ halo.

The \cii\ line detection in J0210-0456 is narrow (FWHM = $189 \pm
18$\,km\,s$^{-1}$) and shows only a small velocity gradient ($\approx
100$\,km\,s$^{-1}$) across a scale of 6\,kpc. The inclination angle of
the \cii\ emission is unconstrained, but the narrow line suggests the dynamical
mass of the system is much lower than would be found in the local
universe for a galaxy hosting a $10^8 \, {\rm M}_\odot$ black
hole. This is in agreement with the results for more massive black
holes at $z\approx 6$ (Wang et al. 2010) and fits with the discussion
above that shows optically-selected $z\approx 6$ quasars have
experienced enhanced black hole accretion relative to their stellar
mass accumulation.

These early observations with ALMA are a prelude to increased resolution and
sensitivity observations to come. A larger sample of $z\approx 6$
quasars with a wide range of black hole masses is necessary to get a more
complete picture of the relationship between black hole growth and
star formation. Spatially resolved observations on scales of $\sim
100$\,pc will reveal the dynamical state of the star forming gas and
enable more accurate determination of dynamical masses.

%% Included in this acknowledgments section are examples of the
%% AASTeX hypertext markup commands. Use \url without the optional [HREF]
%% argument when you want to print the url directly in the text. Otherwise,
%% use either \url or \anchor, with the HREF as the first argument and the
%% text to be printed in the second.

\acknowledgments

Thanks to Javier Graci\'a-Carpio for providing unpublished far-IR data
on the comparison sample of galaxies. Thanks to staff at the North
America ALMA Regional Center for processing the ALMA data. Thanks to
the anonymous referee for suggestions that improved the
manuscript. This paper makes use of the following ALMA data:
ADS/JAO.ALMA\#2011.0.00243.S. ALMA is a partnership of ESO
(representing its member states), NSF (USA) and NINS (Japan), together
with NRC (Canada) and NSC and ASIAA (Taiwan), in cooperation with the
Republic of Chile. The Joint ALMA Observatory is operated by ESO,
AUI/NRAO and NAOJ. The National Radio Astronomy Observatory is a
facility of the National Science Foundation operated under cooperative
agreement by Associated Universities, Inc.


%% in parentheses, after the keyword, but they will not be verified.

{\it Facility:} \facility{ALMA}.

\clearpage

%% the file, uncomment the \plotone, \plottwo, and
%% If you need a layout that cannot be achieved with \plotone or
%% \plottwo, you can invoke the graphicx package directly with the
%% \includegraphics command or use \plotfiddle. For more information,
%% please see the tutorial on "Using Electronic Art with AASTeX" in the
%% documentation section at the AASTeX Web site, http://aastex.aas.org/
%%
%% The examples below also include sample markup for submission of
%% supplemental electronic materials. As always, be sure to check
%% the instructions to authors for the journal you are submitting to
%% for specific submissions guidelines as they vary from

%% This example uses \plotone to include an EPS file scaled to
%% 80% of its natural size with \epsscale. Its caption
%% has been written to indicate that additional figure parts will be
%% available in the electronic journal.

\end{document}